\documentstyle[twocolumn,epsf,aps]{revtex}

                       \def\g{\gamma}        
\def\G{\Gamma}                      \def\D{\Delta}        
           \def\s{\sigma}                 
                             
            \def\r{\rho}                  
\def\k{{\bf k}}            \def\Ek{E_{\bf k}}        
\def\DN{{\cal D}_{\rm S}}  \def\DFs{{{\cal D}_{\rm F}^\s}}              
       \def\S{{\cal S}}                                 
                           
           \def\N{{\cal N}}       \def\Qtot{Q_{\rm tot}}
                                    
\def\lS{l_S}                                                
\def\/{\over}              \def\<{\langle}         \def\>{\rangle}      
\def\us{\uparrow}          \def\ds{\downarrow}     \def\dmu{\delta\mu}  
\def\[{\left[}             \def\]{\right]}                              
\def\({\left(}             \def\){\right)}                              
\def\tT{\tau_{\rm t}}                                  
\def\tS{\tau_{S}}          \def\tQ{\tau_{Q^*}}                          
                                          
\def\GS{\G_{S}}                                          

\def\fks{f_{\k\s}}

\def\TMR{{\D R\/R_F}}                             
             
\def\Pt{{\tilde P}}        \def\Jc{J_{\rm c}}                           
\def\I{I_{\rm inj}}                             
\begin{document}

\title{ \Large \bf   Spin injection and magnetoresistance \\
    in ferromagnet/superconductor/ferromagnet 
    tunnel junctions}   
\author{S. Takahashi, H. Imamura, and S. Maekawa }
\address{Institute for Materials Research, Tohoku University, Sendai 980-8577, Japan}
\maketitle
\widetext
{\begin{abstract}
\rightskip 2.2cm
\leftskip 2.2cm
We theoretically study the spin-dependent transport in a                
ferromagnet/superconductor/ferro-magnet double barrier tunnel junction. 
The spin-polarized tunneling currents give rise to spin imbalance in the
superconductor.  The resulting nonequilibrium spin density suppresses   
the superconductivity with increase of the tunneling currents.          
We focus on the effect of asymmetry in the double tunnel junction,      
where the barrier height of the tunnel junction and the spin-polarization
of the ferromagnets are different, on spin injection, and discuss how   
the superconductivity is suppressed in the asymmetric junction.         
Our results explain recent experimental results on the critical current 
suppression  in high-$T_c$ SCs by spin injection.                       
\end{abstract} }
\rightskip 0cm
\leftskip 0cm


 \topskip 5.0cm
\narrowtext
\bigskip
\bigskip
\bigskip
\medskip

\noindent{\bf I. INTRODUCTION}
\medskip

Spin-polarized tunneling plays an important role in the spin-dependent
transport of magnetic nanostructures \cite{meservey}.  
First the spin-polarized tunneling causes a large magnetoresistance in
ferromagnetic single tunnel junctions  \cite{FF1};
the tunnel resistance decreases when the ferromagnetic moments are
aligned in a magnetic field \cite{FF2}.  
Second the spin-polarized tunneling current driven from ferromagnets (FM)
into a normal metal (N) or a superconductor (SC) creates a nonequilibrium
spin polarization in N or SC \cite{johnson,johnsonS}.  
Recent experiments have shown that a strong suppression of superconductivity
occurs by injection of spin-polarized electrons in tunnel junctions
consisting of a high-$T_c$ superconductor and a ferromagnetic manganite
 \cite{vasko,dong,koller,lee,yeh}.

A double tunnel junction containing SC sandwiched between two FMs
(FM/SC/FM) is a unique system to investigate the nonequilibrium phenomena
of spin and charge imbalance in SC caused by the tunneling currents
but also the competition between superconductivity and magnetism
induced by spin polarization in SC.
In a symmetric double junction, where the tunnel barriers and the
ferromagnets are the same, we have predicted an intriguing magnetoresistive
effect; in the antiferromagnetic ({AF}) alignment of magnetizations,
the spin density accumulated in SC strongly reduces the superconducting
gap $\D$ with increase of tunneling currents, while in the ferromagnetic
({F}) alignment there is no such effect because of the absence of spin
population in SC \cite{FSF}. 
In this paper, we take into account the asymmetry in the junction,
and discuss how the spin density is accumulated in SC and suppress the
superconductivity of SC, depending on the difference in the tunnel
resistance of the barriers and in the spin polarization of FMs.

\bigskip
\noindent{\bf II. FORMULATION}
\medskip

We consider a FM1/SC/FM2 double tunnel junction as shown in Fig.~1.
The left and right electrodes are made of different ferromagnets and
the central one is a superconductor with thickness $d$.
The magnetization of FM1 is chosen to point up and that of FM2 is
either up or down.   In the asymmetric tunnel junction,
the height of the tunnel barriers and/or the strength of the ferromagnets
are different, which are characterized by the different values of the
tunnel resistance and those of the spin-polarization in the junction.

 \topskip 0cm

We calculate the tunneling current using a phenomenological tunneling
Hamiltonian.   If SC is in the superconducting state, it is convenient
to rewrite the electron operators $a_{\k\s}$ in SC in terms of the
quasiparticle operators $\g_{\k\s}$ using the Bogoliubov transformation
  $$
   a_{\k\us} = u_\k\g_{\k\us} + v_\k^*\g^\dagger_{-\k\ds}, \ \ \ \ \
   a^\dagger_{-\k\ds} = - v_\k\g_{\k\us}+u^*_\k\g^\dagger_{-\k\ds} ,
  $$
where $|u_\k|^2 = 1-|v_\k|^2 = {1\/2}\( 1+{\xi_\k/E_\k}\)$ with
the quasiparticle dispersion $E_\k=\sqrt{\xi_\k^2+\D^2}$ of SC,
$\xi_\k$ being the one-electron energy relative to the chemical
potential which is chosen to be zero and $\D$ being the gap parameter.
Then, using the golden rule formula, we obtain the spin-dependent
currents $I_{j\s}$ across the $i$th junction:
\begin{mathletters}                                             
  \begin{eqnarray}                                              
    I_{1\us} &=& \({G_{1\us}/e\DN}\) \[\N_1 - \S - Q^*/2 \] ,   
      \label{eq:I1-u} \\                                        
    I_{1\ds} &=& \({G_{1\ds}/e\DN}\) \[\N_1 + \S - Q^*/2 \] ,   
      \label{eq:I1-d} \\                                        
    I_{2\us} &=& \({G_{2\us}/e\DN}\) \[\N_2 + \S + Q^*/2 \] ,   
      \label{eq:I2-u} \\                                        
    I_{2\ds} &=& \({G_{2\ds}/e\DN}\) \[\N_2 - \S + Q^*/2 \] .   
      \label{eq:I2-d}                                           
  \end{eqnarray}                                                
\end{mathletters}                                               
\begin{figure}
   \epsfxsize=0.8\columnwidth \centerline{\hbox{ \epsffile{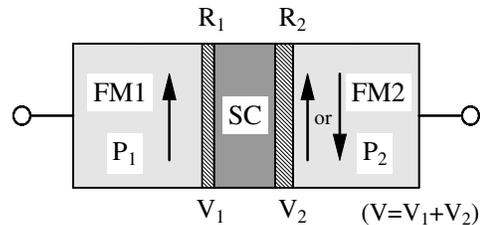}}
     }
\caption{ 
Double barrier tunnel junction consisting of two ferromagnets
(FM1 and FM2) and a superconductor (SC) separated by thin insulating
barriers.  
   }
\end{figure}
\noindent
Here, $G_{i\s}$ ($i=1,2$) is the tunnel conductance of the $i$th
junction for electrons with spin $\s$ if SC is in the normal state,
and is given by $G_{i\s}\propto |T_i^\s|^2\DN \DFs_i$, where
$|T_i^\s|^2$ is the tunneling probability of the $i$th junction
and $\DN$ and $\DFs_i$ are the spin-subband densities of states
in SC and FM$i$, respectively.   The quantity $\N_i$ is given by
\cite{tinkham}
  \begin{equation}                              
    \N_i =  {1\/2} \sum_\k                      
      \Bigl[ f_0\bigl(\Ek-{eV_i}\bigr)          
           - f_0\bigl(\Ek+{eV_i}\bigr)\Bigr] ,  
     \label{eq:N}                               
  \end{equation}                                
where $f_0$ is the Fermi distribution function of thermal equilibrium
in FM and $V_i$ the voltage drop at the $i$th junction ($V_1+V_2=V$).
The quantities $\S$ and $Q^*$ are quasiparticle spin and charge densities
in SC and are defined by
  \begin{eqnarray}                              
    \S = {1\/2} \sum_\k \(f_{\k\us} - f_{\k\ds}\),  \ \ \ \
     Q^* = \sum_{\k\s} q_\k f_{\k\s},           
     \label{eq:SandQ}                           
  \end{eqnarray}                                
where $f_{\k\s}= \<\g^\dagger_{\k\s} \g_{\k\s}\>$ is the distribution
function of quasiparticles with energy $E_\k$ and spin $\s$
and $q_\k=|u_\k|^2-|v_\k|^2$ is the effective charge of a quasiparticle
in the state $\k$.   

The conservation of total charge
$\Qtot=\sum_{\k\s}\<a^\dagger_{\k\s}a_{\k\s}\>$ 
in SC gives $I_{1\us}+I_{1\ds}=I_{2\us}+I_{2\ds}$, which yields the
relation
  \begin{equation}                                      
    \(g_1P_1+g_2\Pt_2\)\S + Q^*/2 = g_1\N_1-g_2\N_2,
     \label{eq:SQ}                                      
  \end{equation}                                        
where $g_i=G_i/(G_1+G_2)$ ($g_1+g_2=1)$ is the reduced
conductance of $i$th junction, and 
  $$                                            
      P_1  = \(G_{1\us}-G_{1\ds}\) / G_1, \ \ \ 
     \Pt_2 = \(G_{2\us}-G_{2\ds}\) / G_2,       
  $$                                            
where $\Pt_2=P_2$ for the {F} alignment and $\Pt_2=-P_2$ for the {AF}
alignment of magnetizations.  
$P_1$ and $P_2$ are the degree of spin-polarization of FM1 and FM2.

The quasiparticle spin density $\S$ generated in SC is calculated by
balancing the spin injection rate
$\(d\S/dt\)_{\rm inj}$ =$[(I_{1\us}-I_{1\ds})-(I_{2\us}-I_{2\ds})]/2e$
with the spin relaxation rate $\S/\tS$, where $\tS$ is
the spin-relaxation time.  
The result is
  \begin{eqnarray}                              
    \S &=& { g_1g_2(P_1 - \Pt_2) \/             
     1 - (g_1P_1+g_2\Pt_2)^2+\GS} \(\N_1+\N_2\),
      \label{eq:S}                              
  \end{eqnarray}                                
where $\GS=g_1g_2(\tT/\tS)$, $\tT=2e^2\DN(R_1+R_2)$ is the
tunneling time, and $R_j=1/G_j$.  Note that the spin density in SC is
proportional to the {\it difference} $(P_1-P_2)$ for the {F} alignment
and the {\it sum} $(P_1+P_2)$ for the {AF} alignment.

The quasiparticle charge density $Q^*$ is obtained by balancing the
injection rate $\(dQ^*/dt\)_{\rm inj}$ with the relaxation rate
$Q^*/\tQ$ \cite{pethick}, where $\tQ$ is the charge relaxation
time \cite{ss}, and using Eq.~(\ref{eq:SQ}) in the form
  \begin{eqnarray}                                              
    Q^* = - \(\tQ\/\tT\) \sum_\k {\D^2\/E_\k^2} \Bigl[ g_1\N_{1\k} - g_2\N_{2\k} 
   \cr
        - (g_1P_1+g_2\Pt_2) \(f_{\k\us}-f_{\k\ds}\)  \Bigr],    
     \label{eq:Q}                                               
  \end{eqnarray}                                                
where $\N_{i\k} = (1/2)[f_0(\Ek- {eV_i})$
 $-f_0(\Ek+{eV_i})] $.

The superconducting gap $\D$ in SC is determined by $\fks$ through
the BCS gap equation 
  \begin{eqnarray}                      
    {1\/V_{\rm BCS}} = \sum_\k          
     {1-f_{\k\us} - f_{\k\ds}\/ E_\k } .
    \label{eq:gap}                      
  \end{eqnarray}                        

It follows from Eqs.~(\ref{eq:S}) and (\ref{eq:Q}) that, if the junction
is symmetric, both $\S$ and $Q^*$ vanish for the {F} alignment, while
$\S \ne 0$ and $Q^*=0$ for the {AF} alignment. 
In the asymmetric case, $\S$ and $Q^*$ become finite for both alignments.
In the following, we restrict ourselves to the case $\tQ \ll \tT \ll \tS$,
where the charge imbalance is very small ($Q^* \sim 0$), so that
the nonequilibrium effect is dominated by the spin imbalance.
In addition, the thickness of SC $d$ is much smaller than the spin
diffusion length $\lS=\sqrt{{D}\tS}$, ${D}$ being the diffusion
constant, so that the distribution of quasiparticles is spatially uniform
in SC.  Then, the distribution function $f_{\k\s}$ is described by $f_0$,
but the chemical potentials of the spin-up and spin-down quasiparticles
are shifted oppositely by $\dmu_S$ from the equilibrium one to generate
the spin density; 
  \begin{equation}                      
     f_{\k\us} = f_0(\Ek - \dmu_S),     
       \ \ \ \                          
     f_{\k\ds} = f_0(\Ek + \dmu_S).     
     \label{eq:fks}                     
  \end{equation}                        
We solve self-consistently Eqs.~(\ref{eq:SandQ}) - (\ref{eq:gap})
with respect to $\D$, $\dmu_S$, and $V_i$, and obtain $\D$ and $\S$
as functions of $V$.  The results are used to calculate the total
tunneling current $\I=I_{i\us}+I_{i\ds}$:
  \begin{eqnarray}                      
    \I = {1\/e\DN} \(\N_1+\N_2 \/ R_1+R_2\)     
    { 1-(g_1P_1^2+g_2P_2^2)   + \GS \/  
      1-(g_1P_1  +g_2\Pt_2)^2 + \GS } , 
     \label{eq:I}                       
  \end{eqnarray}                        
which we call the injection current.

\bigskip
\noindent{\bf III. RESULTS}
\medskip

We briefly discuss the tunnel magnetoresistance (TMR) in the normal
state ($T>T_c$), in which $\N_1+\N_2=\DN eV$, so that the TMR ratio,
$\D R/R_F=(R_A-R_F)/R_F$ has the form
  \begin{eqnarray}                      
    \TMR = { 4g_1g_2P_1P_2              
         \/ 1-(g_1P_1+g_2P_2)^2+\GS}.   
     \label{eq:TMR}                     
  \end{eqnarray}                        
The TMR is degraded in the case of strong asymmetry in the conductances
($G_1 \ll G_2$ or $G_1 \gg G_2$).
A large TMR ratio is obtained when the following conditions are satisfied;
the tunnel barriers are similar ($R_1 \sim R_2$) and the spin relaxation
time in SC is long compared with the tunneling time ($\tT/\tS <1$).
The latter condition is $\(\r_N/R_1\)+\(\r_N/R_2\) > \(Ad/\lS^2\)$, where
$\r_N$ is the resistivity of SC in the normal state and $A$ the junction
area, which requires a low junction resistance and/or a thin
SC with $d$ much smaller than $\lS$.  If these conditions are satisfied,
we have the optimum ratio ${\D R/R_F} \sim P_1P_2 / (1 - P_1P_2)$ in the
normal state.
\begin{figure}
   \epsfxsize=0.95\columnwidth \centerline{\hbox{ \epsffile{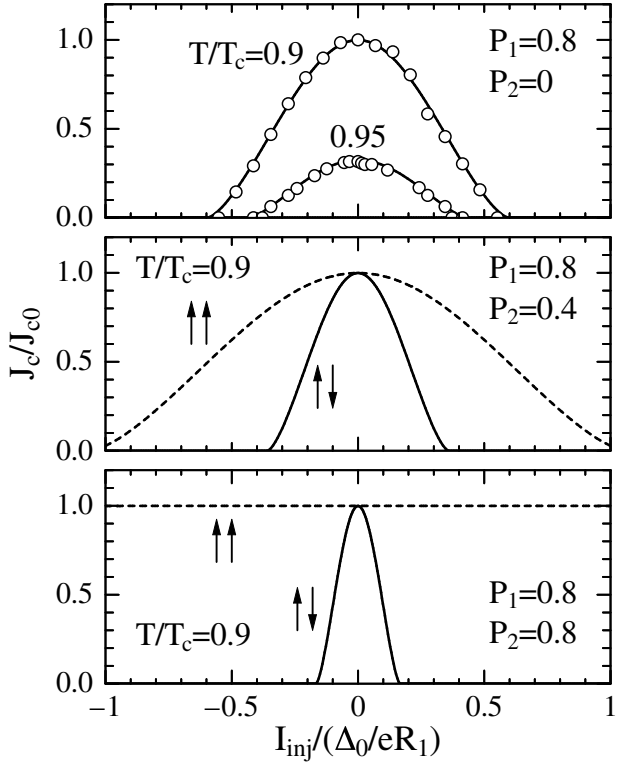}}
     }
\vspace{5pt}
\caption{ 
Dependence of the critical current ${\Jc}$ on the injection current $\I$
for the spin polarization $P_1=0.8$ of FM1 and different values of $P_2$
of FM2.  
The open circles indicate the critical current measured
at $T=80$ K and $T=84$ K ($T_c\sim 89$ K) 
in a FM/SC/N junction made of a high-$T_c$ SC and a ferromagnetic
manganite with $P\sim 100 \%$ \protect\cite{dong}.   
   }
\end{figure}

When temperature $T$ is lowered below $T_c$ at zero bias, SC becomes
superconducting.   As the injection current $\I$ increases,
the superconductivity is strongly suppressed by the pair breaking
effect due to the increase of the quasiparticle spin density $\S$
in SC.
The amount of $\S$ accumulated in SC is directly connected to the
injection current $\I$ by the relation
  \begin{eqnarray}                      
    \S = \[{P_1 -\Pt_2 \/ 1-(g_1P_1^2+g_2P_2^2) + \GS }\]
      { e\DN \/ G_1+G_2 } \I .  
     \label{eq:S-I}                     
  \end{eqnarray}                        
When FM1 and FM2 are the same ($P_1 = P_2$), the injected spins vanish
in the {F} alignment, and are accumulated only in the {AF} alignment.
Therefore the pair breaking effect occurs only in the {AF} alignment.
However, when the FMs are different, the injected spins are accumulated in
proportion to ($P_1 - P_2$) for the {F} alignment and ($P_1 + P_2$) for the
{AF} alignment, and thus we have the pair breaking effect in both alignments.

The suppression of the superconducting gap $\D$ by spin injection is
detected by measuring the critical current $\Jc$.   According to the
Ginzburg-Landau theory, $\Jc$ is proportional to $\D^3$, because
$\Jc \propto \D^2v_c$, $v_c$ being the critical superfluid velocity, and
$v_c \propto \D$ \cite{tinkham}.
Figure 2 shows the cube of the normalized gap, $(\D/\D_0)^3$,
and thus the normalized critical current, $\(\Jc/{\Jc}_0\)$, as a function
of the injection current at temperature $T/T_c=0.9$ for $P_1=0.8$ and three
values of $P_2=$ 0, 0.4, and 0.8.  Other parameters are taken to be
$g_i=1/2$ ($R_1=R_2$) and $\GS=0$.
In the case that FM1 and FM2 are the same ferromagnets (bottom panel), the
critical current $\Jc$ in the {AF} alignment steeply decreases and vanishes at
a small value of $\I$, whereas $\Jc$ in the {F} alignment shows no dependence
on $\I$.  In the case that FM1 and FM2 are different (middle panel),
the critical current decreases with increase of injection current in both
alignments but in different way; $\Jc$ decreases more slowly in the {F}
alignment than in the {AF} alignment.   
If one of the ferromagnets, FM2, is replaced by a normal metal (N), we have
a heterostructure junction FM1/SC/N, which corresponds to the junction
with $P_2=0$ (top panel).   The calculated result for $P_2=0$ explains
the critical current suppression by spin injection observed in the
heterostructure junctions consisting of a high-$T_c$ SC and a ferromagnetic
manganite with $P\sim 100$ [6-10].

\medskip
\bigskip
\noindent{\bf ACKNOWLEDGEMENTS}
\medskip

This work is supported by a Grant-in-Aid for Scientific Research
Priority Area for Ministry of Education, Science and Culture of Japan,
CREST, and NEDO, and by the supercomputing facilities in IMR, 
Tohoku University.

\bigskip



\end{document}